# Superconducting platinum silicide for electron cooling in silicon

M J Prest[1], J S Richardson-Bullock[1], Q T Zhao[2], J T Muhonen[3], D Gunnarsson[4], M Prunnila[4], V A Shah[1], T E Whall[1], E H C Parker[1] and D R Leadley[1]

[1] Department of Physics, University of Warwick, Coventry CV4 7AL, United Kingdom.
[2] Peter Grünberg Institute (PGI 9), Forschungszentrum Jülich, 52425 Jülich, Germany.
[3] Centre for Quantum Computation & Communication Technology, The University of New South Wales, Sydney, New South Wales 2052, Australia.
[4] VTT Technical Research Centre of Finland, P.O. Box 1000, FI-02044 VTT Espoo, Finland.

E-mail: m.j.prest@warwick.ac.uk

We demonstrate electron cooling in silicon using platinum silicide as a superconductor contact to selectively remove the highest energy electrons. The superconducting critical temperature of bulk PtSi is reduced from around 1 K to 0.79 K using a thin film (10 nm) of PtSi, which enhances cooling performance at lower temperatures and enables electron cooling to be demonstrated from 100 mK to 50 mK.

PACS: 72.15.Eb, 63.20.kd, 85.30.De

## 1. Introduction

Schottky junctions form natural tunnel barriers between a degenerate semiconductor and a superconductor [1]. This type of superconducting tunnel contact to an electron gas in a semiconductor can be used as an energy selective filter, removing hot electrons and replacing them with cold electrons [2]. Such junctions can be useful for electron cooling at low temperatures.

The thermal coupling between a crystal lattice and an electron gas, therein, can become very weak at low temperatures (< 1 K) [3]. Hence an electron gas within a normal metal, or degenerately doped semiconductor, can be cooled below the lattice temperature with moderate cooling power.

Provided that the electron and lattice temperatures are sufficiently below the superconducting transition temperature $T_C$ of the superconductor, the cooling power increases with temperature ($T^{3/2}$) [2] but at a lower rate than the lattice heat load (typically $T^5$ or $T^6$) [4, 5]. So in principle, cooling should improve as temperature is reduced; however, at the lowest temperatures other heat loads may become dominant, such as sub-gap leakage currents or ambient heat [6]. The sub-gap leakage can be reduced by using narrow energy gap superconductors, as we have done in this work.

We report here, the application of platinum silicide as the superconductor in a tunnel junction refrigerator. Platinum silicide has a $T_C$ of about 1 K, but this is suppressed in thin films, for layer thicknesses below about 50 nm [7]. The thinnest films also have the best crystallinity, resulting in a smoother surface, and hence a better junction quality. We have used a 10 nm thick PtSi film in our device.

## 2. Experimental

PtSi-Si tunnel junctions were fabricated to form a superconductor – semiconductor – superconductor structure (S-Sm-S). The device cross-section is shown in figure 1a. A silicon on insulator substrate was used and the active areas were ion implanted with arsenic, giving a dopant concentration of $8 \times 10^{19}$ cm$^{-3}$ after activation. Platinum was deposited and silicidation was performed at 500°C for 1 min. Finally, aluminium contacts were patterned by lift-off. Hall bar current-voltage (I-V) measurements gave a sheet resistance of about 100 Ω/square. The sheet resistance and carrier density values agree well with standard curves for silicon [8]. Figure 1b shows the device in plan-view. The implanted region between the contacts is just over 3 times longer than its width (5 μm), giving a series resistance $R_S$ of 320 Ω.

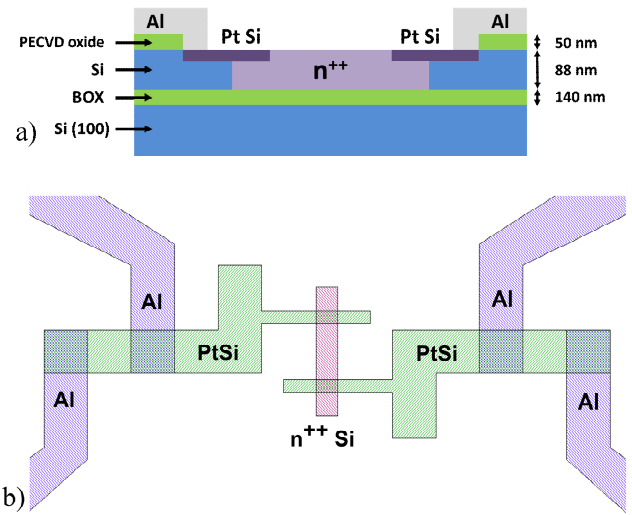

Figure 1. a) Schematic cross-section. A silicon-on-insulator substrate was used with a 140 nm buried silicon dioxide layer (BOX). b) Device layout. The length of the central n++ island is approximately 30 μm and the junction areas are 2.5 μm by 5 μm. The four Al contacts allow a four point measurement, eliminating any voltage drops in the Al-PtSi junctions.

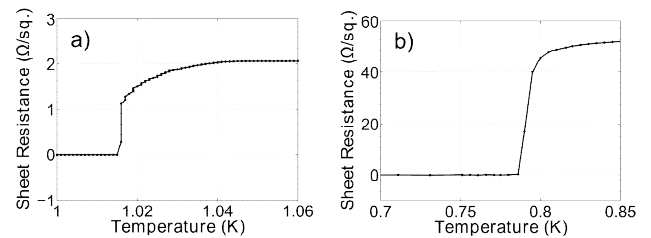

Figure 2. Sheet resistance as a function of temperature, showing superconducting transition temperature $T_C$ of PtSi for layer thicknesses of (a) 100 nm, (b) 10 nm.

In figure 2 we compare $T_C$ measurements for PtSi films of different thickness on low doped silicon substrates. The $T_C$ is reduced from 1.015 K for the 100 nm film, to 0.786 K in our 10 nm film. This reduction seems reasonable, but is a little less than expected (a $T_C$ of ~0.6 K was obtained for a similar film) [7].

*I-V* measurements were performed on the PtSi device at 100 mK using a dilution refrigerator and are shown with a logarithmic current axis in figure 3a. The same data are also shown with a linear current axis (inset) and used to calculate the differential conductance *dI/dV* (figure 3b).

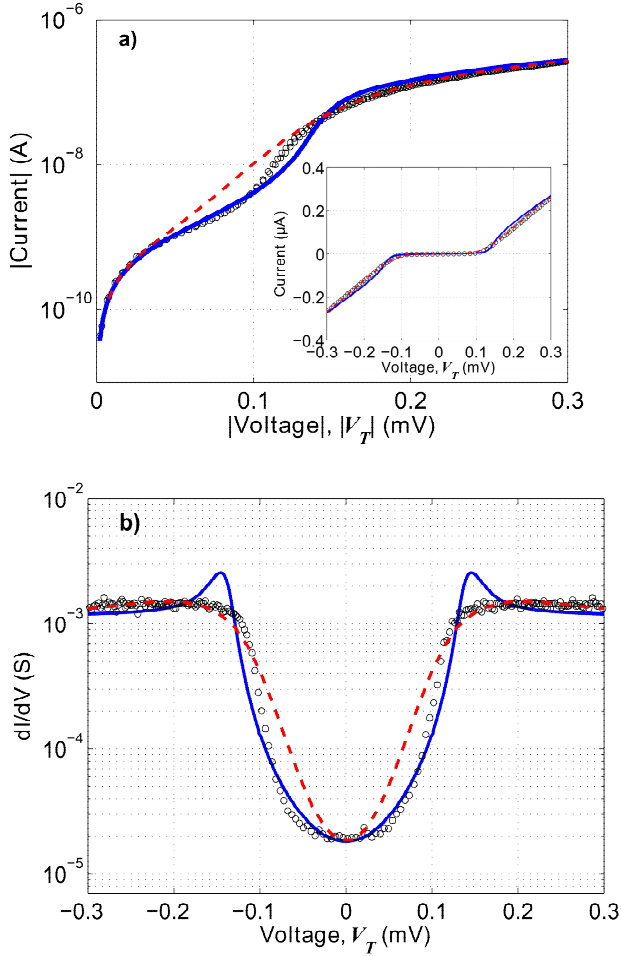

Figure 3. (a) Log current axis *I-V* results for the PtSi cooler (the inset shows the same data with a linear current axis) (b) Differential conductance. All figures use the same experimental data measured at 100 mK (black circles). The dashed curves (red) show the 100 mK isotherm fit to the experimental data. The isotherm model uses the parameters $\Delta = 70$ μeV, $\Gamma/\Delta = 0.8 \times 10^{-2}$, $R_T = 300\ \Omega$, $R_S = 320\ \Omega$. The solid curves (blue) use a cooling model, with additional parameters: $\Sigma = 3.1 \times 10^8$ Wm$^{-3}$K$^{-6}$ and $\nu = 1.41 \times 10^{-17}$ m$^3$.

We have fitted the *I-V* data with electron temperature $T_e$ calculated using [9]

$$I = \frac{1}{2eR_T} \int_{-\infty}^{\infty} \left[ f(E - \tfrac{1}{2}eV_C, T_e) - f(E + \tfrac{1}{2}eV_C, T_e) \right] g(E,\Gamma)\,dE. \qquad (1)$$

where the Fermi distribution function $f(E,T) = 1/[1+\exp(E/k_BT)]$, $k_B$ is the Boltzmann constant, $e$ is the electron charge and the Dynes function for the density of states of a superconductor is given by [10]

$$g(E,\Gamma) = \left| \mathrm{Re}\left[ \frac{E + i\Gamma}{\sqrt{(E + i\Gamma)^2 - \Delta^2}} \right] \right|. \qquad (2)$$

The Dynes leakage parameter $\Gamma$ implies the presence of gap states in the superconductor and is generally used as a figure of merit for superconductor tunnel junctions, with a lower value indicating a better quality junction.

The voltage drop across the PtSi junctions $V_C$ is adjusted to give the total voltage across the device $V_T = V_C + I R_S$.

When solving for $T_e$, (*cooling model*) [11] we use an expression for the cooling power of the S-Sm-S junctions

$$P_c = \frac{2}{e^2 R_T} \int_{-\infty}^{\infty} (E - \tfrac{1}{2}eV_C) \left[ f(E - \tfrac{1}{2}eV_C, T_e) - f(E, T_b) \right] g(E,\Gamma)\,dE \qquad (3)$$

which is used in a heat balance equation

$$P_C + P_{e-ph} + P_J = 0 \qquad (4)$$

and solved for $T_e$. The electron-phonon coupling heat power $P_{e-ph} = \Sigma \nu (T_e^6 - T_b^6)$ where $\Sigma$ is the electron-phonon coupling constant, $\nu$ is the volume of the electron gas and $T_b$ is the bath temperature [12-13]. The Joule heating power $P_J = -I^2 R_s$.

Using (1) and (2) with $T_e = 100$ mK (*isotherm model*) we obtained a reasonable fit to the data as shown by the dashed red curves in figures 3a and 3b. We found the tunnel resistance $R_T = 300\ \Omega$ (3.75 k$\Omega$ μm$^2$), superconductor half-gap $\Delta = 70$ μeV and Dynes leakage parameter $\Gamma/\Delta = 8 \times 10^{-3}$. This value is fairly typical of other semiconductor devices [11], but not as low as that found in normal metal based coolers [6].

In figure 3b, the isotherm (dashed red curve) falls to the minimum with straight sides on the log-linear plot. This exponential behaviour is typical of an isotherm plot and provides a clear distinction from a device with cooling [14]. Figure 3a shows that the isotherm model does not capture the lower current around $V_T = 0.1$ mV, although it fits well for higher biases ($V_T > 0.14$ mV). Reduced current relative to that calculated in the isotherm model is characteristic of cooling in the device [14].

When using (1) to (4) with $T_e$ set by the solution of (4) (*cooling model*), we are better able to reproduce the experimental data in the sub-gap region (for bias $|V_C| < 2\Delta/e$). We used $\Sigma = 3.1 \times 10^8$ Wm$^{-3}$K$^{-6}$ from Kivinen *et al.* [13] for a similar sample and the volume $\nu = 32$ μm $\times$ 5 μm $\times$ 88 nm $= 1.41 \times 10^{-17}$ m$^3$. The model predicts cooling from 100 mK to about 50 mK, as shown in Fig. 4. The cooling power at 100 mK was 0.72 pW (or 29 fWμm$^{-2}$). Also in figure 4, the dashed green curve shows the predicted cooling if all the parameters were the same except for Δ = 190 μeV; as might be expected for a thicker layer of PtSi or aluminium. The larger Δ allows more leakage current and hence prevents cooling below the bath temperature.

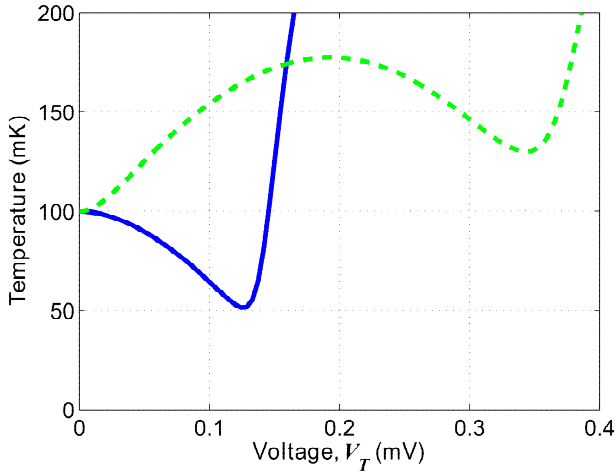

Figure 4. Temperature-Voltage calculations for a bath temperature of 100 mK. Solid curve (blue) uses the cooling model and parameters of figure 3. Dashed curve (green) uses the same parameters except for a larger delta (Δ = 190 μeV).

Small discrepancies in the experimental and predicted data are most noticeable at biases beyond sub-gap ($|V_C| > 2\Delta/e$) and are highlighted by the sensitivity of the $dI/dV$ curve. As shown in figure 3b, the cooling model predicts peaks in $dI/dV$ at the ± 2Δ/e bias points; however, these peaks are very weak in both the experimental data and the isotherm model; also, the $I$-$V$ curves display an abrupt transition from sub-gap to normal state resistance at these points (as shown in the inset of figure 3a).

By inspection of (1), we see that $dI/dV$ should have maxima which correspond to peaks in the superconductor density of states (2). The peak height is reduced in the presence of sub-gap leakage (we have $\Gamma/\Delta = 8\times10^{-3}$ which is quite high compared to conventional NIS junctions) [6]. Also, high series resistance spreads the voltage dependence. These factors (high sub-gap leakage and high series resistance in our device) explain why the peaks are diminished for the experimental data and the isotherm model.

For the cooling model, we suggest that the $dI/dV$ peaks are enhanced by junction heating for biases beyond 2Δ/e. Perhaps, in the measured device, the junction heating is not as strong as predicted by the cooling model, either because of damping due to heat dissipation in the device, or perhaps the density of states of PtSi is not as strongly peaked as predicted by the Dynes expression (2), although the mid-gap density of states is consistent with the experimental results. Quasiparticle recombination and back tunnelling could be included to improve the model [15].

## 3. Conclusions

PtSi is known to act as a superconductor, and because silicides have been widely used as contact materials in the semiconductor industry (due to their reliability and good electrical characteristics) it seemed natural to try PtSi for electron cooling in silicon. We have fabricated the first PtSi/Si/PtSi electron cooling device and shown that when PtSi is used as a thin layer (10 nm), its $T_C$ is reduced (and superconducting gap 2Δ is narrowed) which is beneficial for cooling at low bath temperatures. As there is a significant sub-gap leakage current in our device ($\Gamma/\Delta = 8\times10^{-3}$), the tunnelling of low energy electrons could reduce the cooling power. By narrowing the gap, we reduce the sub-gap leakage current, and this aids in providing greater effective cooling power. A calculation of cooling using a larger Δ (figure 4, dashed green curve), with other parameters the same as those used to fit the experimental data, demonstrates that sub-gap leakage prevents cooling in that case; this highlights the advantage of the thin layer with low Δ in our sample.

Experimental data are compared to a cooling model (with varying electron temperature as predicted by an energy balance equation) and an isotherm model (with constant electron temperature). The experimental data are best described by the cooling model, particularly for biases $V_T < 0.1$ mV, where most of the cooling occurs. The fit provides good evidence for electron cooling from 100 mK to 50 mK. Slight discrepancies between the model and the data are more marked at higher biases. The abrupt transition from sub-gap to normal state resistance, in the experimental data, deviates from the cooling model, so may be a characteristic of the materials or geometry of our device.

The tunnel barrier to holes of PtSi is lower than that of aluminium, so PtSi could also be useful for a hole cooler. It has recently been shown that holes have a higher thermal coupling to the lattice than electrons [16], so require more power to cool; however, for applications that involve cooling of the lattice, or sensing of lattice temperature, PtSi could be beneficial.

## Acknowledgements

This work has been financially supported by EPSRC through Grant No. EP/F040784/1 and by EC through Project 257375 Nanofunction Network of Excellence and by the Academy of Finland.